# Robust N-1 secure HV grid flexibility estimation for TSO-DSO coordinated congestion management with deep reinforcement learning

Zhenqi Wang[1]*, Sebastian Wende-von Berg[1,2], and Martin Braun[1,2]
[1]University of Kassel, Kassel, Germany
[2]Fraunhofer Institute for Energy Economics and Energy System Technology (IEE), Kassel, Germany
* zhenqi.wang@uni-kassel.de

*Abstract* - Nowadays, the PQ flexibility from the distributed energy resources (DERs) in the high voltage (HV) grids plays a more critical and significant role in grid congestion management in TSO grids. This work proposed a multi-stage deep reinforcement learning approach to estimate the PQ flexibility (PQ area) at the TSO-DSO interfaces and identifies the DER PQ setpoints for each operating point in a way, that DERs in the meshed HV grid can be coordinated to offer flexibility for the transmission grid. In the estimation process, we consider the steady-state grid limits and the robustness in the resulting voltage profile against uncertainties and the N-1 security criterion regarding thermal line loading, essential for real-life grid operational planning applications. Using deep reinforcement learning (DRL) for PQ flexibility estimation is the first of its kind. Furthermore, our approach of considering N-1 security criterion for meshed grids and robustness against uncertainty directly in the optimization tasks offers a new perspective besides the common relaxation schema in finding a solution with mathematical optimal power flow (OPF). Finally, significant improvements in the computational efficiency in estimation PQ area are the highlights of the proposed method.

I. INTRODUCTION

The coordination between TSO and DSO is essential in efficiently solving the grid congestions caused by the massive DER integration within the last two decades. The flexibility from the integrated DERs is needed, on the one hand, for internal redispatch measures in the distribution grids and, on the other hand, to communicate possible available flexibility to the TSO.

Multiple latest works have investigated the application of mathematical methods in estimating PQ flexibility from controllable assets, e.g., DER, On-load tap changer (OLTC), Static Synchronous Compensator (STATCOM) at the TSO-DSO interface. J. Silva et al. proposed a nonlinear Optimal Power Flow (OPF)-based interval constrained power flow algorithm, which solves a set of optimization problems to characterize the flexibility boundary with consideration of the technical constraints and illustrate the PQ area with the cost of flexibility [1]. L. Lopez, et al. proposed and investigated multiple constraints relaxation schemes (convexation/linearization) to improve the computational efficiency with minor sacrifice on the estimation accuracy [2]. D. A. Contreras, et al. further considers the technical constraints of DER according to the German grid code, while the optimization problem is completely linearized to improve the computational efficiency and the convergence [3]. N. Savvopoulos, et al. proposed an estimation method focusing on the residential PVs and battery storage system with an optimization-based method [4]. M. Sarstedt, et al. presented an optimization-based method with the population-based aggregation methods to find PQ area with cost optimal flexibility demand to single DER [5].

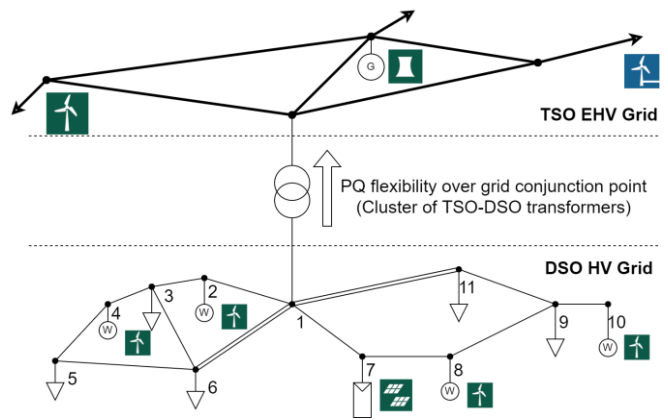

Figure 1 The concept of estimating TSO-DSO PQ flexibility of HV grid.

In this work, the objective is to estimate the PQ area at the TSO-DSO interface of HV grids. The large area HV grid type is the study focus, which, as shown in Figure 1, has a meshed topology. The most reviewed works focus on the DSO grid with none or low meshed topology, such as the MV grids. The N-1 criterion is a major challenge for the OPF. The criterion means that the outage of a single HV line should not lead to an overloading of the rest of the system. To consider this in the mathematical optimization is still challenging so far. A feasible approach was developed in project EU-SysFlex [6], where the OPF-calculations are performed on the base case (N-0 case) as well as N-1 cases, which inevitably leads to higher computational overhead.

Another major challenge is handling the underlying uncertainties from the forecasted grid status. Specifically, for the operative planning, the uncertainties from the wind and solar forecasting and the aggregated HV/MV load can lead to errors in estimating the PQ area since some setpoints will result in a high probability of violating grid constraints.

To overcome these challenges, our work aims to develop a deep reinforcement learning (DRL)-based method to estimate the PQ flexibility from DERs in meshed HV grids at the TSO-DSO interfaces, which is the first of its kind. The HV grid security needs to be fulfilled, especially focusing on the above-mentioned N-1 and robustness against probable voltage violations caused by uncertainties, which is a significant improvement of the state-of-the-art research works. By using Artificial Neural Network (ANN) to predict DER setpoints for PQ flexibility estimation, high computational efficiency can be achieved.

This paper is organized as follows. Section 2 introduces the proposed ANN-OPF method for deterministic N-0 secure PQ flexibility estimation. Section 3 extends the method with two supervised trained approximators to estimate N-1 criterion and voltage violation probability under uncertainties and to use them for reinforcement training. In section 4, the proposed methodology is verified with case studies and results, regarding the approximation quality, prediction quality, and computational efficiency.

## II. DETERMINISTIC N-0 SECURE FLEXIBILITY ESTIMATION WITH ANN-OPF

DRL is a subcategory in machine learning in which an agent (an ANN) is to be self-supervised trained to fulfill tasks through its interaction (denoted as reward/penalty) with a training environment. In [7], a method called ANN-OPF is proposed based on this concept to carry out standard OPF tasks self-supervised trained with a fast parallel power flow calculation (PFC) solver [8]. The PFC solver can solve 10,000 PFCs of, e.g., IEEE case118 in under 0.1s. Thus, it is efficient enough to be used as the training backend. This PFC solver evaluates an augmented loss function and computes the action gradients from each decision variable. The augmented loss function consists of an objective term and penalty terms from constraints violation. The method can be combined with supervised training to improve the optimality of predicted setpoints. After the training step, the ANN can predict optimized setpoints upon the observed grid status and further inputs during the application step.

In this section, the PQ flexibility estimation with deterministic voltage profile and N-0 secure constraints are modeled mathematically, and the ANN-OPF will be extended to fulfill this functionality.

### A. Mathematical formulation

The TSO-DSO PQ flexibility estimation can be mathematically formulated as follows, where B, E, D, L, G and T denote all the buses, branches, controllable DERs, loads, generators, and TSO-DSO transformers, respectively. The objective of the PQ flexibility estimation is to minimize the deviation of $P_i^t, Q_i^t$ with the optimized DER setpoints $P_i^{DER}, Q_i^{DER}$ to the PQ requirements at the TSO-DSO transformer $P_i^{t,sp}, Q_i^{t,sp}$.

$$\underset{P_i^{DER}, Q_i^{DER}}{\text{minimize}} \sum_{i \in \mathbf{T}} (|P_i^t - P_i^{t,sp}| + |Q_i^t - Q_i^{t,sp}|)$$

subject to

$$P_i + Q_i \cdot j = \sum_{k \in \mathbf{B}} \underline{V} \cdot (\underline{Y}_{(i,k)}^{bus} \cdot \underline{V}_k)^*, \quad \forall i \in \mathbf{B},$$

$$V_i^{\min} \leq V_i \leq V_i^{\max}, \quad \forall i \in \mathbf{B},$$

$$I_i^{side} = |\underline{Y}_{(i,b_f)}^{side} \cdot \underline{V}_{b_f} + \underline{Y}_{(i,b_t)}^{side} \cdot \underline{V}_{b_t}|, \quad \forall i \in \mathbf{E}, \forall side \in (f,t),$$

$$lp_i = \frac{\max(I_i^f, I_i^t)}{I_i^{\max}} \cdot 100\%, \quad \forall i \in \mathbf{E},$$

$$lp_i \leq lp_i^{\max}, \quad \forall i \in \mathbf{E},$$

$$P_i = P_i^G + P_i^{DER} - P_i^L, \quad \forall i \in \mathbf{B},$$

$$Q_i = Q_i^G + Q_i^{DER} - Q_i^L, \quad \forall i \in \mathbf{B},$$

$$0 \leq P_i^{DER} \leq P_i^{DER,\max}, \quad \forall i \in \mathbf{D},$$

$$Q_i^{DER,\min,P^{DER}} \leq Q_i^{DER} \leq Q_i^{DER,\max,P^{DER}}, \quad \forall i \in \mathbf{D}$$

For deterministic N-0 secure PQ flexibility estimation, the constraints formulation follows the convention of the standard AC-OPF, where the equality constraints of power flow equations, the inequality constraints of the voltage profile limit on $V$, and branch loading limits $lp$ are required.

Furthermore, the HV DER also needs to be modeled and formulated as required in VDE-AR-N 4120 [9], especially the PQ working area of DER (exemplarily shown in Figure 2). Unlike the stationary limits, the available Q range ($Q^{DER,\min,P^{DER}}, Q^{DER,\max,P^{DER}}$) of DER is related to the P operating point. Especially for the area where the P operating point drops below 20% of the installed capacity of a DER. The Q range is strongly affected and, in this case, reduced.

The equality constraints of power flow equations are enforced directly through the PFC and the technical requirements on DER are enforced with the DER modeling. The inequality constraints on the deterministic voltage profile violation and N-0 branch loading violation need to be solved with the predicted DER-setpoints and named "**hard constraints**".

### B. Augmented loss function

To train the ANN to fulfill the designated task, an augmented loss function needs to be formulated from the mathematical formulation of the optimization problem. First, the objective term needs to be normalized as follows.

$$\underset{P_i^{DER}, Q_i^{DER}}{\text{minimize}} \sum_{i \in \mathbf{T}} \left( \frac{|P_i^t - P_i^{t,sp}|}{P_i^{t,\max} - P_i^{t,\min}} + \frac{|Q_i^t - Q_i^{t,sp}|}{Q_i^{t,\max} - Q_i^{t,\min}} \right)$$

The $P_i^{t,\max}, P_i^{t,\min}, Q_i^{t,\max}, Q_i^{t,\min}$ are calculated based on the input grid status, such as the load PQ and DER $P^{DER,\max}$ manually adapting the DER setpoints $P^{DER}, Q^{DER}$ to its maximum and minimum. This process will only be required once in the sample creation. This step is essential to avoid overweighting of the objective term over the penalty term. The penalty term for hard constraints violations is defined as follows.

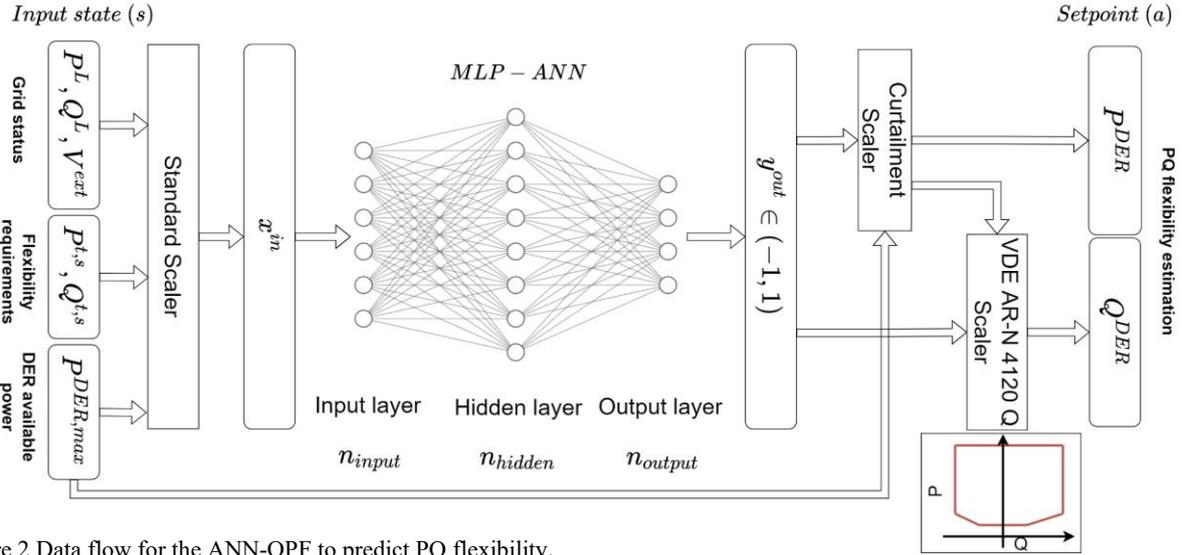

Figure 2 Data flow for the ANN-OPF to predict PQ flexibility.

$$l_V = \sum_{i=1}^{B} \max(V_i^{\min} - V_i, V_i - V_i^{\max}, 0)$$

$$l_{lp} = \sum_{i=1}^{E} \max(lp_i - lp_i^{\max}, 0)$$

The violations are penalized with constant penalization factors, which are 100, and 1 for voltage profile violation $\omega_V$ and branch loading violation $\omega_{lp}$, respectively, which are identical to the configuration used in [7]. The resulting augmented loss function for training is formulated as follows.

$$l_{aug} = obj_{flex} + w_V \cdot l_V + w_{lp} \cdot l_{lp}$$

### C. Data flow of ANN-OPF

Figure 2 depicts the data flow of the ANN-OPF for PQ flexibility estimation. The grid status vector is required as input, which consists of load PQ, DER available power $\text{P}^{\text{DER,max}}$, and voltage at the external grid. The PQ flexibility requirements at the TSO-DSO transformer are also required. All the features will be concatenated and standardized.

The output layer of the ANN is a "tanh" function, the output of which ranges in (-1, 1). Thus, the output needs to be converted to fulfill the technical requirements of DER. The P of DERs is limited by the minimum of its natural availability and the installed capacity defined as $\text{P}^{\text{DER,max}}$. For this purpose, a curtailment scaler is needed to scale the ANN output to the range of $(0, \text{P}^{\text{DER,max}})$. The Q availability is limited in this case by the size of the installed capacity as well as the current P operational point. A so-called VDE-AR-N 4120 converter is implemented to perform the P-related Q scaling.

### III. ROBUST N-1 SECURE FLEXIBILITY ESTIMATION

So far, the ANN-OPF can predict deterministic N-0 secure PQ flexibility. This section proposes approximation methods to predict N-1 constraints violations and voltage profile violation probabilities, called "**soft constraints**". Consequently, the extension of ANN-OPF integrating the approximators in training for robust and N-1 secure PQ flexibility estimation is introduced.

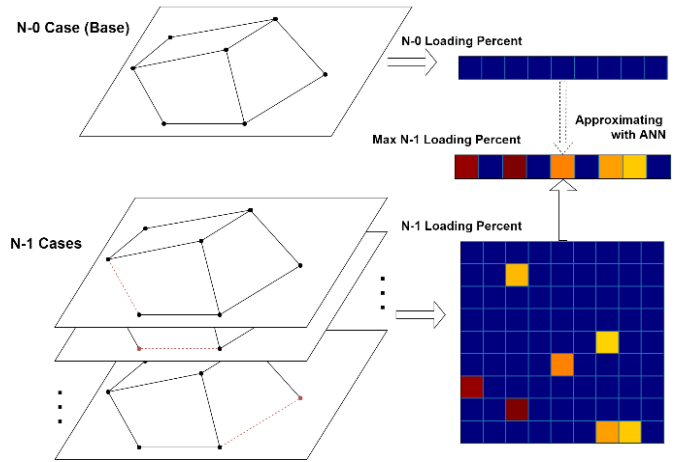

Figure 3 Concept of N-1 analysis.

### A. N-1 approximation

To consider N-1 constraints with ANN-OPF, the N-1 constraint violation needs to be penalized during the training process. A complete N-1 evaluation can be carried out as follows. As shown in Figure 3, besides performing PFC on the base case, the PFCs need to be performed on all N-1 cases. In this work, the N-1 cases consider only the line outage. The results of the N-1 analysis are the max. line loading among all the N-1 cases, which is defined as $lp_{n1}$. This process is possible with the fast parallel PFC-solver. However, during the numerical gradient evaluation for the training process, such computational overhead will be unacceptable. A computational efficient supervised training-based approach to approximate the identification of N-1 violations is proposed.

Multiple supervised training-based machine learning methods for fast N-1 approximation are proposed and evaluated in [10] using the infeed and load characteristics as input features. For the ANN-OPF training, instead, we use the PFC results of the base case with the purpose of improving the approximation quality.

The input feature for the ANN approximator are the PFC results of the base case, such as line loading and voltage profiles. The combinations of the features are studied in detail. The output feature is then the $lp_{n1}$. For supervised training, the correct labels are required. In this case, a complete N-1 calculation on the sample will be carried out with the fast parallel PFC-solver. The loss function "smooth L1 loss" is utilized for training, with which the large deviations are penalized with quadratic function while the slight deviations are penalized with a linear function.

### B. Probabilistic power flow approximation

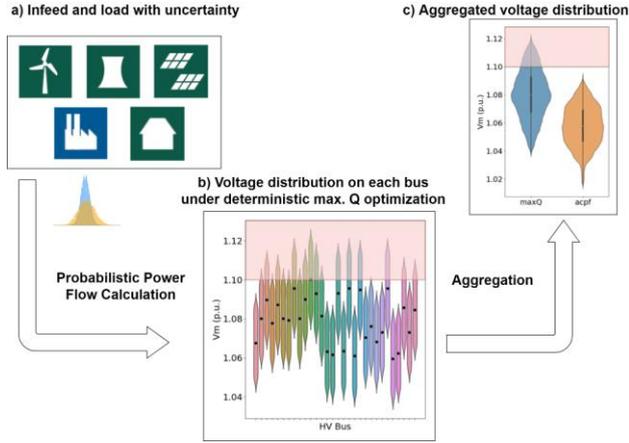

Figure 4 Concept of probabilistic power flow calculation.

Figure 4 shows the concept of using a Monte-Carlo Simulation (MCS)-probabilistic power flow (PPF) to estimate the impact of uncertainties from infeed- and load-forecasting (Figure 4a) on the grid voltage profile of each bus (Figure 4b). This result can be aggregated into an overall voltage profile distribution of the HV level (Figure 4c). In this example as shown in Figure 4b, the results were achieved with a probabilistic evaluation with deterministic max. Q flexibility optimization. It can be observed that multiple buses have a voltage profile violation probability of around 50%, while the overall HV voltage profile violation probability is about 14%. Looking at the reference case with no max. Q optimization (acpf), there is only a small probability for any voltage profile violations under consideration of these uncertainties. Knowing the voltage profile violation probability on each bus is thus essential to improve the robustness of the optimization. Therefore, a supervised learning-based PPF-approximator is proposed so that a complete PPF can be avoided during the training phase.

Besides the grid modeling and the deterministic profiles for grid status, the uncertainty assumptions of the profiles are required for the reference MCS-PPF in label creation. The objective of the PPF-approximator is to predict the voltage violation probability on each bus, which is also the result of the reference MCS-PPF. As the input feature, the results from the deterministic PFCs are used. Here, especially the deterministic voltage profile as well as the bus/DER PQ injection are used. Their combinations are investigated in detail in this case study. In this work, the voltage profile violation probability on single buses under 10% is considered secure.

### C. Overall ANN-OPF training process

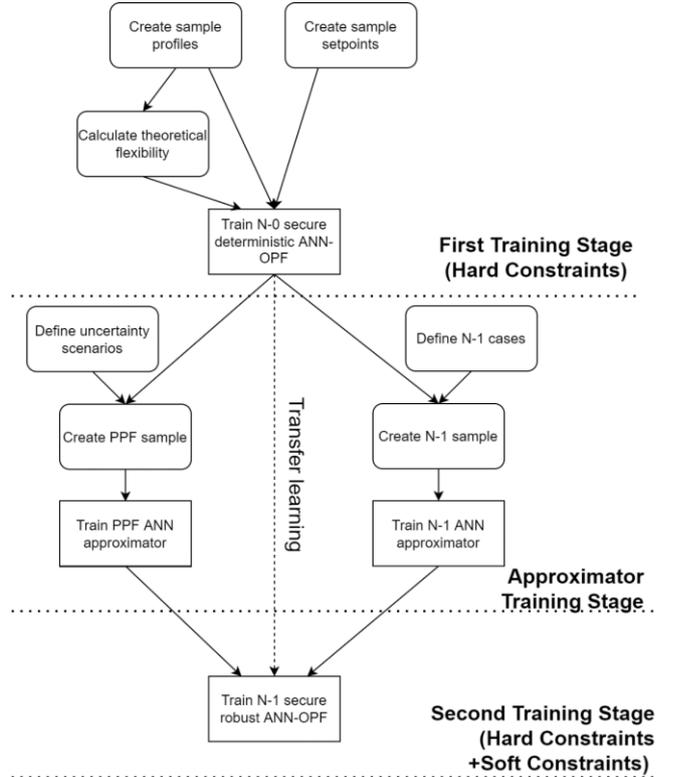

Figure 5 Overall process of training ANN-OPF.

The overall training process of using ANN-OPF to perform robust N-1 secure flexibility estimation is shown in Figure 5 and can be understood as follows. Firstly, the training sample must be created in two steps, which are the generation of the sample profiles considering the correlation of load/DER characteristics and the generation of the PQ requirements sample according to an equal distribution in the range of (0, 1). For the normalized objective evaluation, the theoretical flexibility of each sample needs to be computed. The created sampling is used for the following ANN-OPF and approximator training phases.

In the first stage training, the ANN-OPF is self-supervised trained to find the PQ setpoints for controllable DERs. With the grid status and PQ requirements at the TSO-DSO transformer as input, the ANN-OPF predicts PQ setpoints considering hard constraints.

In the approximator training stage, the fast parallel PFC-solver is used to perform complete N-1 analysis and MCS-PPF for label generation. In this stage, the DER setpoints from the first stage ANN-OPF are required to perform the so-called "on-sample" sample creation. As "off-sample" sample creation, the noised time series data with no DER setpoints is used. The comparison of both sample creation methods is evaluated in the case study.

In the second stage training, a transfer learning process is applied, in which the ANN-OPF from the first stage training will be further self-supervised trained with the additional penalty terms from the approximators. The method is inspired by the approach utilized in robust mathematical optimization as proposed in [11]. The deterministic optimization will be extended to robust optimization with linearized improvements based on the sensitivity to the potential violations under uncertainties.

The training method is realized as follows. The approximator will be applied to identify soft constraints violation from the batched PFC results. The soft constraints violation will be marked for the gradient calculation for the extended augmented loss function.

For instance, the N-1 approximator sees line 5 with N-0 loading of 72% but with 105% N-1 loading. Given all the other lines are not expecting N-1 loading violations, line 5 will be marked in the PFC results from the gradient calculation. The gradient from the batched PFC can be used to improve the soft constraint violation indirectly. Precisely for this case, the action that reduces the N-0 loading on line 5 can be securely used to reduce the N-1 overloading. For the voltage profile robustness training, the bus with a voltage violation probability over 10% will be marked and improved during training.

After this stage, the ANN can learn a decision strategy for predicting setpoints, which reduces the chances of violating the N-1 criterion and the probability of violating voltage limits under uncertainties.

### D. Postprocessing in the application

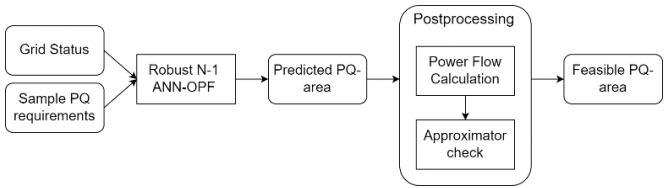

Figure 6 Process of postprocessing in PQ flexibility estimation.

Figure 6 shows to use of the trained ANN-OPF to predict feasible PQ area. Upon the input grid status and PQ requirements at the TSO-DSO transformers, the ANN predicts DER setpoints upon the requirement. To estimate a PQ area, the PQ requirements are generated equally distributed. Although ANN-OPF is trained with penalty from the hard and soft constraints violations, there is no guarantee that the resulting setpoints are feasible. A postprocessing step is required to filter out only the feasible setpoints. In this step, the PFC is performed with the predicted setpoints, which identifies all the hard constraint violations. With the PFC results, the approximators will identify soft constraints violations. The maximum feasible PQ area can be identified with the process.

## IV. CASE STUDY

### A. Test case definition

For the evaluation, we use the open-dataset SimBench HV grid [12] "1-HV-urban1-no_sw". The meshed mid-size grid contains 100 buses. All 22 wind parks in the HV grid are considered controllable. The installed capacity of these wind parks is doubled. The remaining 79 DERs model the aggregated DER from the MV grid, thus assumed as uncontrollable. The possible N-1 cases are defined as the lines whose outage will not lead to unsupplied grid regions. There are in total 78 N-1 cases in the grid. The uncertainties from forecasting are assumed as normal distributed, and its standard deviation $\delta$ is proportional to the absolute forecasted value. The uncertainty assumption on each data points is given as 10% of the current P- and Q values and 1% of the external grid voltage. The default tap position of OLTC is set to 0. The flexibility is considered as the sum of flexibility of the three parallel transformers.

For the ANN-OPF and the approximators, the same ANN structure used in [7] is utilized, where the size of the input layer and output layer equals the input and output feature size, respectively. For the approximators, the hidden layer units are set as 300. 500 hidden layer units are used for the ANN-OPF model, which are identified suitable from multiple tests.

### B. Approximator performance evaluation

Table 1 N-1 approximation results.

| Input feature | $(P,Q)$ | $lp$ | $(lp,V)$ | $(V,\delta)$ |
|---|---|---|---|---|
| | Approx. difference of N-1 line loading (%) | | | |
| Off-sample | 248.67 | 40.6 | 69.35 | 99.84 |
| On-sample | 6.65 | **6.33** | 6.33 | 14.23 |
| | Success rate in N-1 congestion classification (%) | | | |
| Off-sample | 62.61 | 97.5 | 97.34 | 91.65 |
| On-sample | 99.52 | 99.34 | **99.74** | 98.97 |

Table 2 PPF approximation results.

| Input feature | $(P,Q)$ | $V$ | $(V,Q^{DER})$ | $(V,\delta)$ |
|---|---|---|---|---|
| | Approx. difference of voltage violation probability (%) | | | |
| Off-sample | 36879.21 | 38.22 | 39.9 | 28.81 |
| On-sample | 14.25 | 6.54 | **6.27** | 6.41 |
| | Success rate in voltage violation prob. classification % | | | |
| Off-sample | 51.97 | 97.88 | 98.04 | 98.01 |
| On-sample | 97.4 | **98.29** | 98.26 | 98.26 |

The case study of the approximators focuses on two aspects. Firstly, the best feature combination from the base PFC results needs to be identified so that only the features with high relevance are used as input features (the definitions are given in section 2.A). Secondly, to achieve good approximation in the application, the ANN should be trained on data sets that are similar to the data sets for the application since ANN are generally good at interpolation, not extrapolation. The approximation results will be unpredictable if the training data sets and applying data sets are from different distributions. The approximation results are verified with the second stage ANN-OPF considering the "off-sample" and "on-sample" cases comparing to the label from complete N-1 and PPF analysis.

For both approximators, it can be observed from the results in Table 1 and Table 2 that the on-sample training is essential in guaranteeing the approximation quality, which also justifies the proposed overall training process. For N-1 approximation and voltage profile violation probability approximation, it can be observed that the $lp$ and $V$ are the most important features respectively. For the simplicity of the input feature, the approximators with only these features are selected for the training process.

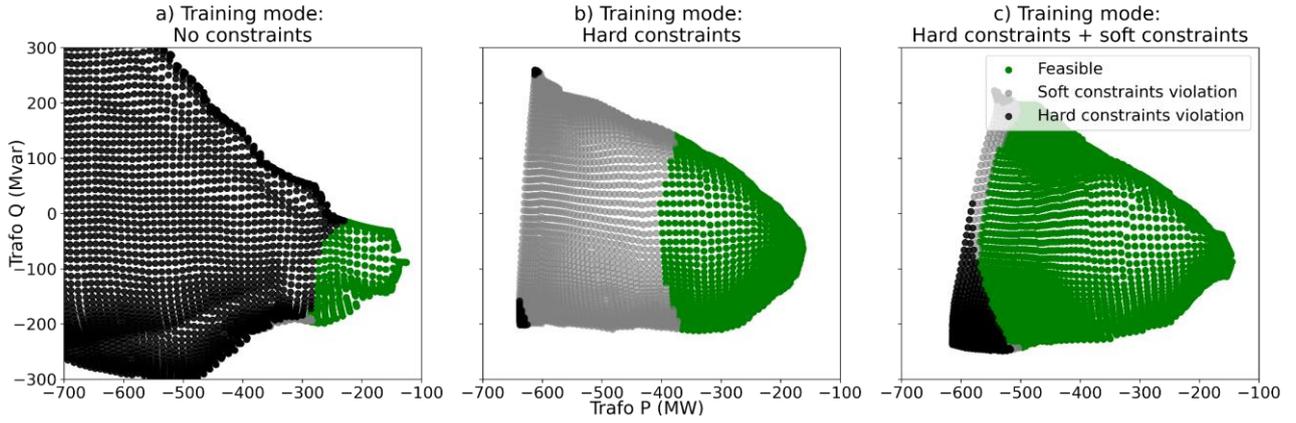

Figure 7 Resulting PQ area of one time step with different training configurations.

*C. Flexibility optimality evaluation*

The ANN-OPF trained with hard constraints is compared against the mathematical OPF pypower PIPS [13] with the pandapower interface [14]. Since the PIPS does not support the PQ flexibility estimation objective and the uncertainty robust/N-1 constraints, this evaluation focused on the max./min. feasible PQ flexibility, where an adapted cost function for PIPS is used. The P cost of DER is set to -1 so that the cost optimization objective maximizes the infeed without causing grid congestions. Analog to the P cost, Q cost is set to -1 and 1 for max. and min. Q optimization respectively. The VDE-AR-N 4120 PQ area is considered statically, which is accomplished by first identifying the max. feasible P and computing Q range based the P setpoints of each DERs. The result is shown under an appropriate OLTC tap position selection of 12.

Figure 8 shows the comparison result. As can be observed, for the max. P comparison, the ANN-OPF follows closely the results from PIPS. In this case, active power curtailment is required to avoid N-0 violations. In the max./min. Q comparison, the ANN-OPF exceeds the performance of the PIPS in most of the time steps. This is caused by the different objective formulation (cost of P/Q) for the PIPS optimization. Overall, it can be summarized that properly trained ANN-OPF can reach or exceed an interior point optimizer. It should be noticed that the mathematical OPF fails to converge on one time step, which is another advantage of the proposed ANN-OPF approach.

*D. Resulting PQ area*

In Figure 7, the results of ANN-OPF predicted PQ area with different penalty configurations during training are shown. The figure shows the predicted PQ area and the violations identified with the postprocessing step. The hard constraints violation and soft constraints are marked with the color black and grey, respectively. Figure 7a shows the PQ area from ANN-OPF trained with no penalization from constraint violation, while Figure 7b and Figure 7c are the trained ANN-OPF from the first stage and second stage training, respectively. The tap position 12 is also used so that max. Q flexibility can be achieved.

Figure 7a gives the largest PQ area since the ANN-OPF tries only to fulfill the PQ requirements at the transformers without considering the grid constraints. As a result, the trained ANN-OPF often predicts setpoints, which are not feasible in reality. Looking at the Figure 7b, where the ANN-OPF is trained with the penalization from hard constraints, the PQ area with hard constraints violation is reduced significantly. Thus, the ANN-OPF from the first stage is suitable for estimating PQ area where only the hard constraints must be considered. The Figure 7c gives the result trained considering both hard and soft constraints, where the maximum feasible PQ area is identified. The significant increase of the feasible PQ area shows the effectiveness of the integration of supervised trained soft constraints approximator into the training process.

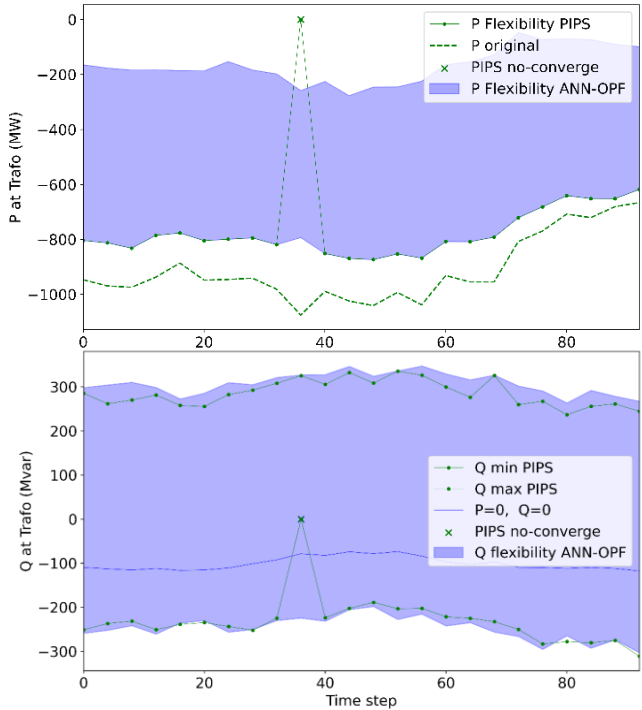

Figure 8 Comparison of ANN-OPF against PIPS optimization (upper: P comparison, lower: Q comparison).

Also, the results show the necessity of the postprocessing step since the ANN-OPF cannot directly predict feasible PQ area due to the complexity of the optimization problem and large decision space.

*E. Computational performance evaluation*

Table 3 Computational efficiency for PQ area prediction.

| Number of samplings | Time prediction (ms) | Time postprocessing (ms) | Total time (ms) | Time 1 OPF (ms) |
|---|---|---|---|---|
| 20x20 | 8 | 26 | 34 | 0.085 |
| 50x50 | 42 | 104 | 146 | 0.058 |
| 100x100 | 91 | 404 | 495 | 0.050 |
| PIPS (1 OPF) | - | - | - | 850 |

Table 3 shows the timing of applying the proposed method. It can be clearly observed that the different numbers of total points are required to estimate the PQ area. The prediction and postprocessing phase can be accomplished significantly below 1 second. The increment of the points does not lead to an enormous increment in total time. At the same time, the PIPS optimization requires the computation of one OPF 850ms, while the ANN-OPF requires for one OPF less than 0.1ms, which shows an acceleration effect of factor x10,000. This computational performance matches the requirement from the real-life grid operation perfectly. It can be used to repetitively predict PQ area when later and better forecasting are available.

## V. CONCLUSION

This work presents a novel deep reinforcement learning method called ANN-OPF to estimate TSO-DSO PQ flexibility. The proposed method considers the N-1 criterion and the robustness of voltage profile under uncertainties by applying the supervised trained ANN approximators. Considering these criterions for PQ flexibility estimation is generally difficult for mathematical OPF. As a result, the feasibility of the predicted PQ area considering the grid constraints can be assured through the postprocessing step and meanwhile delivers high computational efficiency. As next step, the applicability of the proposed method will be further investigated considering aspects such as topology changes.

ACKNOWLEDGMENT

The work was funded within the project "SPANNeND" (FKZ:03EI4040C) of the research initiative "OptiNet" funded by the Federal Ministry of Economics and Energy according to a decision of the German Federal Parliament. Part of the development was funded by the Hessian Ministry of Higher Education, Research, Science and the Arts, Germany through the K-ES project under reference number: 511/17.001 and by the European Union's Horizon 2020 research and innovation programme within the project EU-SysFlex under grant agreement No. 773505. The authors are solely responsible for the content of the publication.